\documentclass[final,5p,times,twocolumn]{elsarticle}  % <-to test final format (2 columns)

\usepackage{hyperref}
\journal{Journal of \LaTeX\ Templates}

\usepackage[referable]{threeparttablex}		%para as foottables com hiperlink
\usepackage{tabularx,booktabs}  % Combines tabularx and longtable functionality

\usepackage{gensymb}
\usepackage{amsmath}
\usepackage{amssymb}

\begin{document}

\begin{frontmatter}

\title{A Large-Area RPC Detector for Muography}
%\title{Elsevier \LaTeX\ template\tnoteref{mytitlenote}}


\author[mymainaddress]{J. Saraiva\corref{mycorrespondingauthor}}

\cortext[mycorrespondingauthor]{Corresponding author}
\ead{joao.saraiva@coimbra.lip.pt}

\author[mysecondaryaddress]{C. Alemparte}
\author[mysecondaryaddress]{D. Belver}
\author[mymainaddress]{A. Blanco}
\author[mysecondaryaddress]{J. Call\'{o}n}
\author[mysecondaryaddress]{J. Collazo}
\author[mysecondaryaddress]{A. Iglesias}
\author[mymainaddress]{L. Lopes}

\address[mymainaddress]{LIP - Laboratory of Instrumentation and Experimental Particle Physics, 3004-516 Coimbra, Portugal}
%\address[mymainaddress]{LIP - Laborat\'{o}rio de Instrumenta\c{c}\~{a}o e F\'{i}sica Experimental de Part\'{i}culas, 3004-516 Coimbra, Portugal}

\address[mysecondaryaddress]{Hidronav Technologies SL, 36202 Vigo, Pontevedra, Spain}

\begin{abstract}

A muon telescope equipped with four Resistive Plate Chambers of 2 m$^{2}$ per plane was tested with the muon scattering tomography technique. 
%A muon telescope equipped with four Resistive Plate Chambers of 2 m$^{2}$ per plane was built and tested at the Laboratory of Instrumentation and Physics in Coimbra, Portugal. In view of a possible application for muon scattering tomography, the telescope was tested during several hours with high atomic number materials located at the center of the telescope with two detectors on each side.
The telescope was operated during several hours with high atomic number materials located at its center with two detector planes on each side. With an intrinsic efficiency above 98\%, spatial resolution around 1 cm and detector planes spaced by 45 cm, it was possible to identify the presence of a 5 cm thick tungsten block in 10 minutes of acquisition. The results obtained after five hours of acquisition are also presented in this communication.
\end{abstract}


\end{frontmatter}

%\linenumbers

\section{Introduction}

The muon scattering tomography, herein referred to as Scattering Muography (SM), is an imaging technique using the highly penetrating capability of cosmic-ray muons to scan, during a relatively short period of time, mid-sized objects with volumes in the order of few to tens of cubic meters. The SM technique finds applications in fields like homeland security and border control, searching for instance for smuggled fissile materials or arms in commercial trucks or sea containers.
%short period of time // small amount of time

Resistive Plate Chambers (RPCs) could be a good choice for the SM technique since they can cover wide areas with high efficiency, spatial and time resolutions, and can be built at moderate cost when compared to other technologies.
%Gaseous detectors such as Resistive Plate Chambers (RPCs) could be a good choice for the SM technique since they can be built at moderate cost when compared to other technologies, covering wide areas with high efficiency, spatial and time resolutions.
%Gaseous detectors such as Resistive Plate Chambers (RPCs) can cover wide areas at moderate cost, with high efficiency, spatial and time resolutions. They could, therefore, be well suited for the SM technique.

In view of a possible application using the SM technique, a telescope of four RPCs was tested during several hours with two detector planes on each side of high atomic number materials.
With an efficiency above 98\%, spatial resolution around 1 cm and detector planes spaced by 45 cm, it was possible to identify the presence of a 5 cm thick tungsten block in 10 minutes of acquisition, as shown later in this paper.

Designed to work outside the traditional laboratory environment and at very low gas flow regime, the telescope was afterwards sent to an industrial complex in Spain.
Some features of the specific design of the telescope include for instance: (1) each detector plane and the respective instrumentation are confined in an external aluminium enclosure (see figure~\ref{rpc}) requiring only three connections to the outside to be operated: gas, power and communication, making maintenance and possible detector replacements in the industrial complex, easier; (2) the active volume around the glass stack in each RPC is encapsulated in a tight polypropylene box, which has excellent water vapor blocking properties as well as good impermeability to atmospheric gases, allowing the operation of the telescope, in stable conditions, with very low gas flow.

\section{Scattering Muography (SM)}
Muons are the most abundant energetic charged Secondary Cosmic Rays (SCR) at sea level and have very high penetration capabilities since they are least likely
to interact with significant loss of energy. With an average momentum of 4 GeV/c at ground level, muons can easily penetrate several meters of rock. These characteristics qualify muons as the best particle to investigate internal properties of large-scale objects.

The Scattering Muography (SM) rely on the Coulomb scattering of muons to infer the presence of materials with high atomic number in large volumes.
\subsection{Multiple Coulomb Scattering (MCS)}
The statistical cumulative effect of many small-angle scatterings, suffered by charged particles traversing a material, is a deflection from their original direction, which angular distribution roughly follows a normal distribution with non-Gaussian tails due to the rare large-scattering events. The Lynch \& Dahl formula (equation~\ref{eq-sigmaDeflection}) provides the rms width of a Gaussian approximation of the MCS angular distribution projected onto a plane containing the initial muon trajectory.~\cite{PDG_2020}

\begin{equation} \label{eq-sigmaDeflection}
\theta_0 = \frac{13.6 \hspace{0.1cm}\mathrm{MeV}}{\beta cp}z\sqrt{\frac{x}{X_0}} \left[1 + 0.038ln \left( \frac{xz^2}{X_0\beta^2} \right)  \right]  \hspace{0.5cm} \mathrm{rad}
\end{equation}

With, $\theta_0$ the rms width of the projected angular distribution provided in radians, $\beta$c the velocity, p the momentum and z the charge number of the incident particle and x/X the thickness of the scattering medium expressed in radiation lengths.

The scattering medium dependence of equation~\ref{eq-sigmaDeflection} is provided by the radiation length (X$_0$) which decreases with the increase of the material's atomic number (Z). Since the width of the MCS angular distribution increases with Z, it is possible to distinguish between high and low-Z materials measuring muon trajectories upstream and downstream a probed volume, as reported for the first time in 2003~\cite{Borozdin_2003}.

Several reconstruction techniques might be used to identify the location of higher deflections in a 3D region. The Point of Closest Approach was the one used with these reported tests.

\subsection{Point of Closest Approach (PoCA)}

\begin{figure}[ht] %[htbp]
\centering
\includegraphics[width=.48\textwidth, trim = 0 0 0 0, clip=true, angle=0]{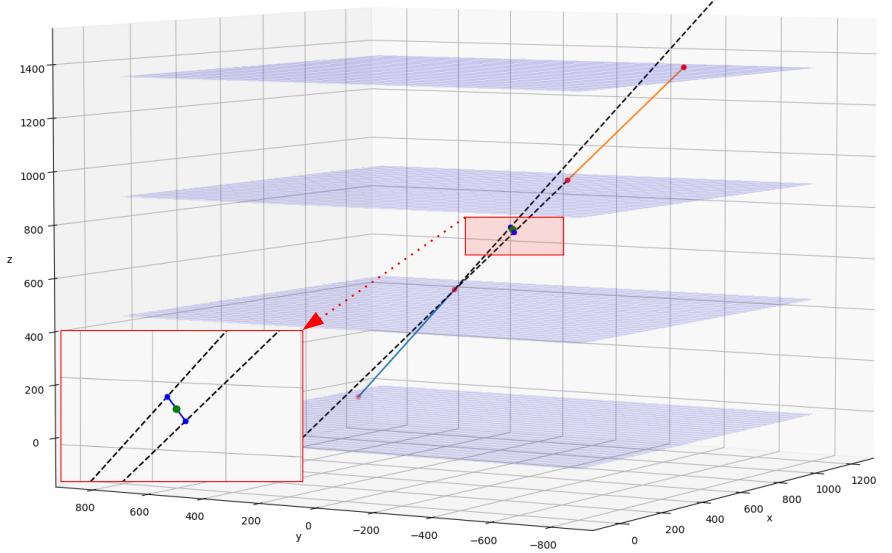}
\caption{Four points of a muon track measured by RPCs (red points in the horizontal planes) obtained when a target of Tungsten (Z=74) was located between the inner detectors; the points in the two upper planes define the incoming trajectory (upstream the target), while the ones in the lower planes define the trajectory downstream the target. The PoCA point is shown in the lower left corner of the figure (green point).}
\label{poca}
\end{figure}

The Point of Closest Approach (PoCA) algorithm can be used to identify the location of high scattering events and in this way, reconstruct a 3D image of the respective object~\cite{Riggi_2013}.
The muon tracks must be measured before and after traversing the target, with detectors located in both sides.
%To differentiate between high and low-Z materials, muon trajectories must be known before and after impinging the object being studied.
It can be done, measuring at least four points of each muon path, two upstream and two downstream the target, as shown in figure~\ref{poca}.

In the case of coplanar incoming and outgoing trajectories, the PoCA point is simply their intersection, otherwise it corresponds to the midpoint of the shortest distance between the two trajectories. When using a large number of trajectories, the algorithm allows the 3D reconstruction of high-Z materials, with each PoCA point from large angular deflections, corresponding to a point in the material.

\section{Muon Telescope}

\begin{figure}[ht] %[htbp]
\centering
\includegraphics[width=.4\textwidth, trim = 0 0 0 0, clip=true, angle=0]{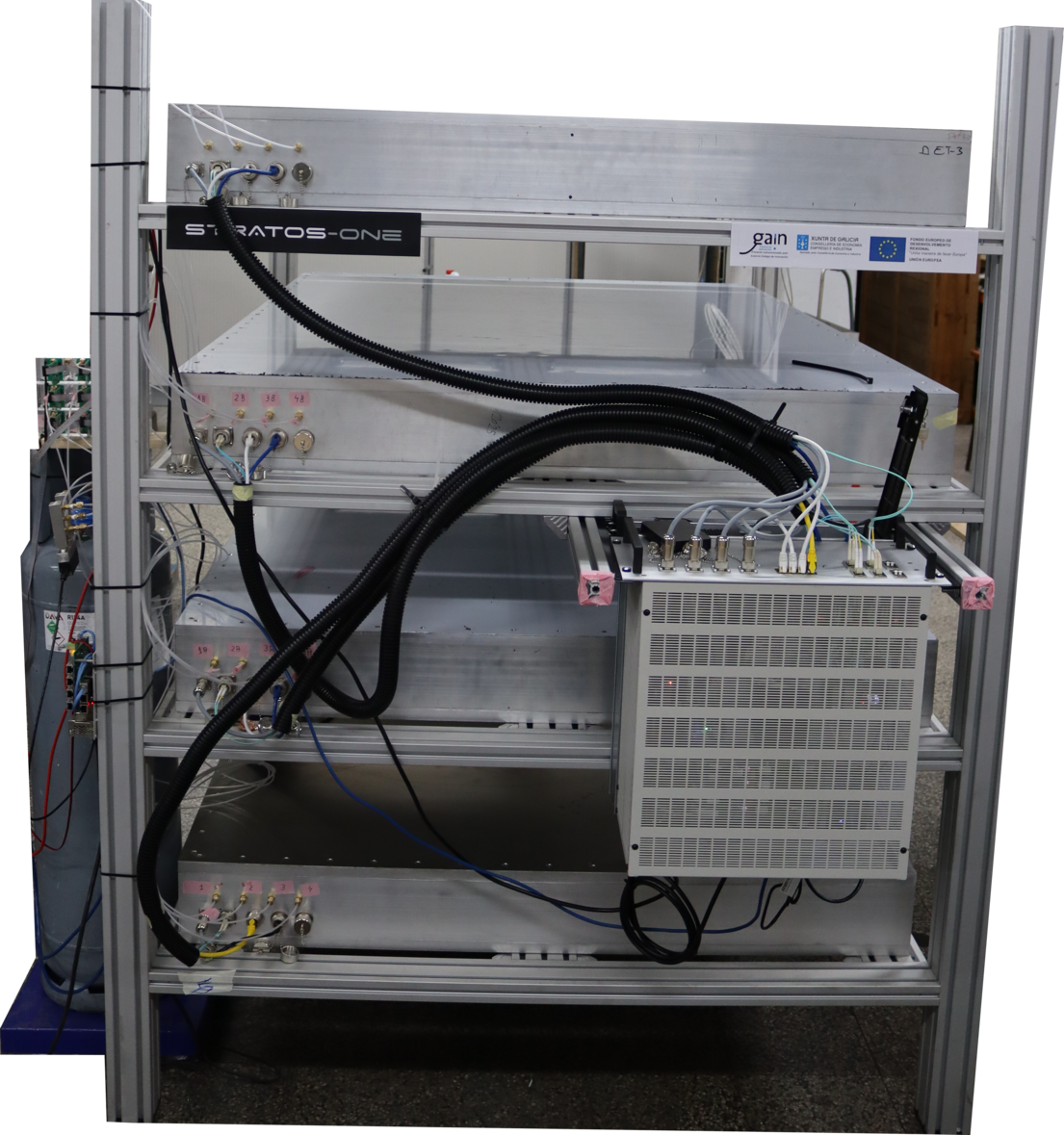}
\caption{Muon telescope of four RPC planes. Each detector and its instrumentation are confined in an aluminium enclosure to avoid external electromagnetic interferences. In the foreground of the image, the DAQ system which includes the Central Trigger Sistem (CTS) and a single board computer running the slow control software and performing the data logging and data storage.
%In the foreground of the image, the DAQ system which includes: the logic and trigger units, readout, event builder, slow control, data logging and data storage.
}
\label{telescope}
\end{figure}

\begin{figure}[h!] %[htbp]
\centering
\includegraphics[width=.35\textwidth, trim = 0 0 0 0, clip=true, angle=0]{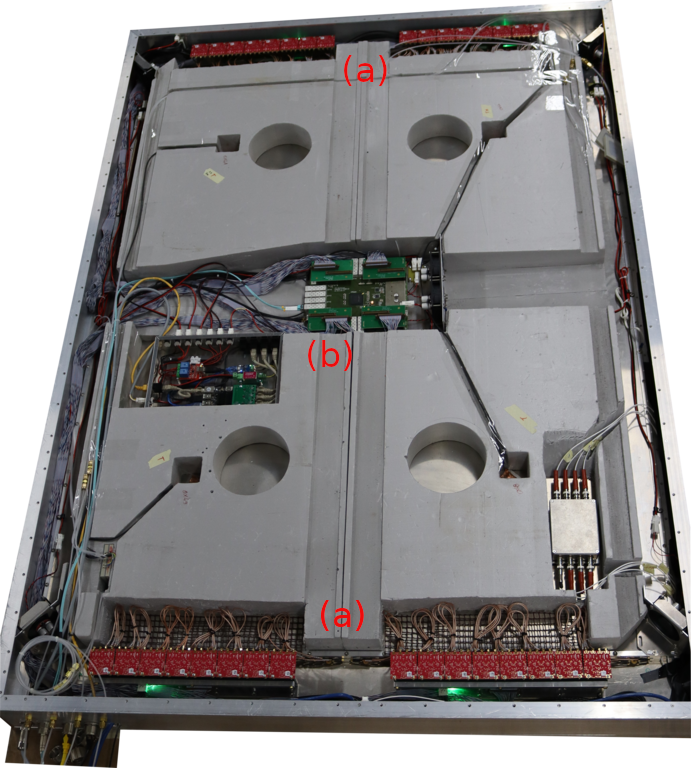}
\caption{One RPC and the respective instrumentation; (a) fast front end electronics with 64 channels at both sides of the RPC; (b) Trigger Readout Board (TRB)~\cite{Neiser_2013} implementing the first level of logic as described below.}
\label{rpc}
\end{figure}

The muon telescope consists of four horizontally oriented RPCs spaced by 45 cm, each with an active area of 1.6$\times$1.2 m$^{2}$ and operated in open gas loop with 1,1,1,2-tetrafluoroethane (R134a), only. The whole setup, including four RPCs and the Data Acquisition (DAQ) System, can be seen in figure~\ref{telescope} while figure~\ref{rpc} shows an RPC and its instrumentation all confined in an external aluminium enclosure. The mentioned distance between planes was used for the sake of convenience to perform the preliminary tests described in this communication. With this configuration, the angular acceptance\footnote{half-angle of a vertically oriented cone.}  of the telescope was approximately 42$\degree$ in one dimension and 50$\degree$ in the orthogonal one (see table~\ref{tab-specifications_Stratos}).
%As a result of the mentioned distance between planes, the angular acceptance of the telescope was approximately 42$\degree$ in one dimension and 50$\degree$ in the orthogonal one (see table~\ref{tab-specifications_Stratos}).

\begin{figure}[ht] %[htbp]
\centering
\includegraphics[width=.46\textwidth, trim = 0 0 0 0, clip=true, angle=0]{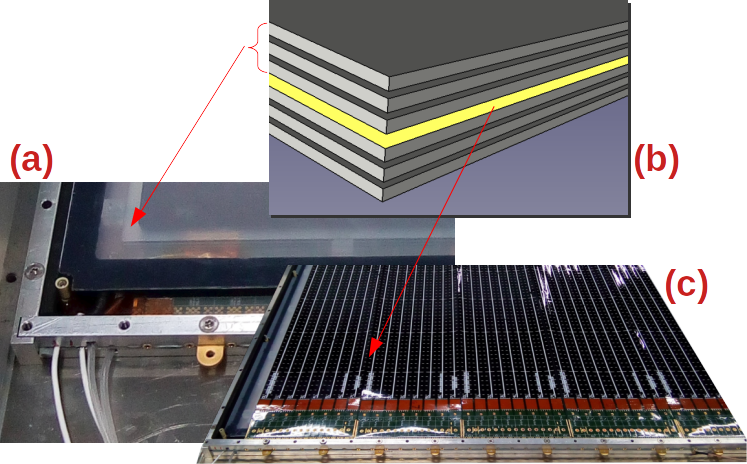}
\caption{(a) RPC module which consists of a sealed plastic (PP) box enclosing the gas volume along with the resistive and HV electrodes. In turn, an internal aluminium case encloses the two modules with the readout electrodes in the middle (lid of the metal housing removed for the picture); (b) layer diagram showing two sensitive volumes, with two gas gaps each, separated by the readout PCB which location is represented by the yellow color (the sealed plastic boxes around each module are not shown for better understanding); (c) PCB with 64 longitudinal signal pick-up strips.}
\label{PPmodules}
\end{figure}

Each RPC is composed of 2 modules with 2 gas gaps each, 1 mm wide, and 3 glass electrodes 2 mm thick with bulk resistivity of $\approx$ 4x10$^{12}$~$\Omega$cm at $25$~$^\circ$C. The readout strips are located between the two modules (see figure~\ref{PPmodules}): a flexible printed circuit board (PCB) of 64 longitudinal signal pick-up strips, 1.55 cm wide and a pitch of 1.85 cm. The gas volume of each module is confined within a polypropylene (PP) box, which includes the resistive electrodes and the high voltage (HV) electrodes (semi-conductive layer, based on an acrylic paint with approximately $100~M\varOmega/\Box$, applied to the outer surface of the outermost resistive electrodes with airbrush technique).

The approach of using sealed plastic boxes made of PP around the sensitive volumes allowed us to operate an RPC in stable conditions with very low gas flow: one volume exchange each 1.5 days for a gas volume of 4200 cm$^{3}$ per RPC module, as described here~\cite{Saraiva_2022}. Additional details regarding the DAQ system and the front-end electronics (FEE) can also be obtained in the provided reference.
%Additional details regarding the DAQ system and the front-end electronics (FEE) are also mentioned in the provided reference.
%Additional details regarding the DAQ system and the front-end electronics (FEE) can also be found in the mentioned reference.

\section{Muon Location in Space and Time}

The signal induced by muons crossing the detector planes are collected by the 64 longitudinal readout strips and fed to pre-amplifiers located at both sides of each RPC (see figure~\ref{rpc}).
In case of pulses with amplitude above a certain threshold, their arrival time and integrated charge are encoded, respectively, by the rising edge and width of a digital signal created by the FEE and sent to the DAQ.
To infer that an RPC detected a particle, a first level of logic requires having, at the same time, a pulse on both sides of at least one readout strip. Additionally, in case of detection by a majority of three out of four RPCs, the event is recorded. 
Of these, only events seen by the four RPCs are of interest for the tests of scattering muography here reported, since the muon trajectory must be known before and after traversing the target, with two RPCs used to define each of these directions, as described further below.

The FEE connected to both sides of each strip provides four measurements: two of time (T$_\mathrm{F}$ and T$_\mathrm{B}$) and two of charge (Q$_\mathrm{F}$ and Q$_\mathrm{B}$), with the subscripts 'F' and 'B' referring to the front and back sides of the detector. The average front and back values: (T$_\mathrm{F}$+T$_\mathrm{B}$)/2 and (Q$_\mathrm{F}$+Q$_\mathrm{B}$)/2, correspond to the time and charge of the induced signal in each strip, respectively. Moreover, to determine the 2D position of a particle in the RPC, the following approach is used:
the strip with the highest value of charge provides the position in one direction (X), while (T$_\mathrm{F}$-T$_\mathrm{B}$)/2 gives the position in the transversal direction, i.e. along the strip (Y).
%the value (T$_\mathrm{F}$-T$_\mathrm{B}$)/2 gives the position along the strip (Y), while the strip with the highest value of charge provides the position in the transversal direction (X).
%Moreover, to determine the 2D position of the particle in each RPC, the following approach is used: the strip with the highest value of charge provides the position in one dimension (X), while (T$_\mathrm{F}$-T$_\mathrm{B}$)/2 gives the position in the transversal one (Y) (along the strip).

\section{Detector Performance}

The main features of the telescope and respective RPCs are summarized in table~\ref{tab-specifications_Stratos} and described below.

\begin{table}[ht]
\centering
\begin{threeparttable}
\begin{footnotesize}
\begin{tabular}{@{}m{4.0cm}m{3cm}@{}}\toprule[1pt]
Time Resolution ($\sigma_t$)
& 300 ps $< \sigma_t <$ 400 ps \\
\midrule[0.1pt]
Spatial Resolution ($\sigma_{x,y}$)
& $\sigma_{x}$: 0.85 cm, $\sigma_{y}$: 1.6 cm \\
\midrule[0.1pt]
Angular Resolution ($\sigma_{\theta_x,\theta_y}$) %\tnotex{angRes-Tnote:1}
& $\sigma_{\theta_x}$: 1.5$\degree$, $\sigma_{\theta_y}$: 2.9$\degree$ \\
\midrule[0.1pt]
Angular Acceptance ($\Omega$)
& 0.5$\pi$ sr $< \Omega <$ 0.7$\pi$ sr \\
\midrule[0.1pt]
Efficiency ($\epsilon$)\tnotex{eff-Tnote:2}
& $\epsilon >$98\% \\
\bottomrule[1pt]
\end{tabular}
\end{footnotesize}
\begin{scriptsize}
\begin{tablenotes} 
%\item[$\ast$] \label{angRes-Tnote:1} tan$^{-1}$(FWHM/(L/2)); L: distance between inner RPCs.
\item[$\ast$] \label{eff-Tnote:2} with a Reduced electric field of 235 Td.
\end{tablenotes}
\end{scriptsize}
\end{threeparttable}
\caption[RPC properties]{Four-plane telescope properties.}
\label{tab-specifications_Stratos}
\end{table}
%the values are provided from the bottom (RPC\_\#1) to the top (RPC\_\#4); the significantly lower efficiency of peripheral RPCs is explained by the inclined muon tracks (high zenith angle) impinging only three of four RPCs

\begin{description}
\item[Time Resolution:] estimated by the difference in time measured by two RPCs; values in the order of few hundreds of picoseconds ($\sigma$) were obtained after applying the time-walk correction;
\item[Spatial Resolution:] achieved by first calculating the residuals of the linear fits to the muon tracks, and applying afterwards a conversion factor based on a quick algorithm generating the normal distributions needed to get the observed residuals;
%achieved by first calculating the residuals of the fits of the muon tracks and the standard deviation of the respective distributions (one per RPC plane and direction (x \& y)). Afterwards, using a quick algorithm generating four normal distributions of random numbers, one for each RPC plane, it was possible to determine which standard deviations ($\sigma_x$ \& $\sigma_y$) the distributions should have in order to obtain the observed spread in the residual distributions;
\item[Angular Resolution:] standard deviation of the angular distribution obtained for vertical muons, with two detectors spaced by 45 cm and with the spatial resolution provided in table~\ref{tab-specifications_Stratos};
\item[Angular Acceptance:] solid angle covered by the telescope and defined by the detector plane dimensions (x: 120 cm; y: 160 cm) and the distance between the outer planes (z: 1350 cm); 
\item[Efficiency:] ratio of the number of events seen by all the RPCs and by only three of them, excluding the one which efficiency is being measured.
\end{description}

\section{Method}

\begin{figure}[ht] %[htbp]
\centering
\includegraphics[width=.45\textwidth, trim = 0 0 0 0, clip=true, angle=0]{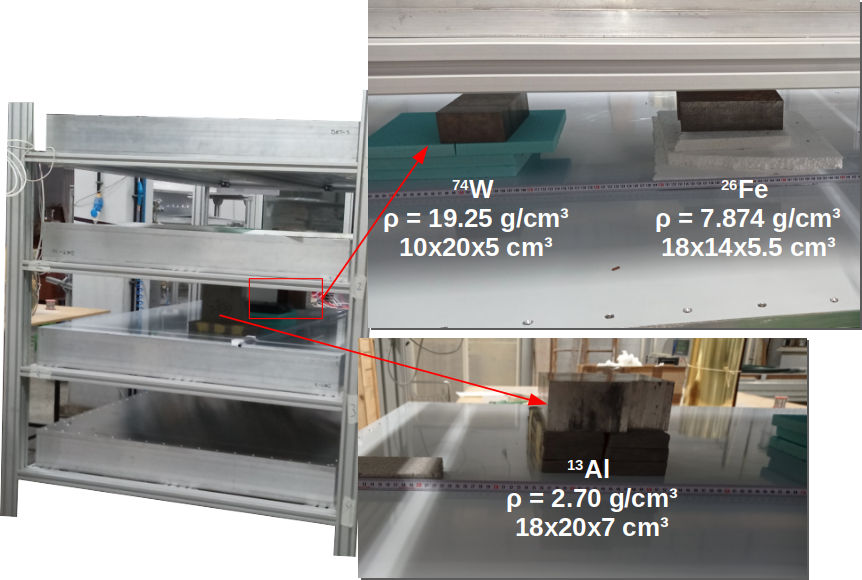}
\caption{Several targets being imaged by the four-plane telescope, namely: aluminium (Al), iron (Fe) and tungsten (W).}
\label{Al_Fe_W_test}
\end{figure}

Preliminary tests of SM were performed for the first time in Coimbra in 2021. Several materials were placed at the center of the telescope as shown in figure~\ref{Al_Fe_W_test}.

After having four intersection points between the muon trajectory and the RPC planes, the following quantities were computed for each event: (1) the direction of the incoming and outgoing muon paths; (2) the angular deflection suffered by the muon; (3) the respective PoCA point.

Since the PoCA algorithm operates on the assumption that the PoCA points correspond to single Coulomb scattering events occurred in the object, its 3D reconstruction can be achieved by plotting all the PoCA points complying a certain angular restriction, as shown in the next section.
%Since the PoCA algorithm operates on the assumption that the PoCA points correspond to single Coulomb scattering events occurred in the object, the 3D reconstruction of the latter can be achieved by plotting all the PoCA points complying a certain angular restriction, as shown in the next section.

\section{Results}

\begin{figure}[ht] %[htbp]
\centering
\includegraphics[width=.45\textwidth, trim = 0 0 0 0, clip=true, angle=0]{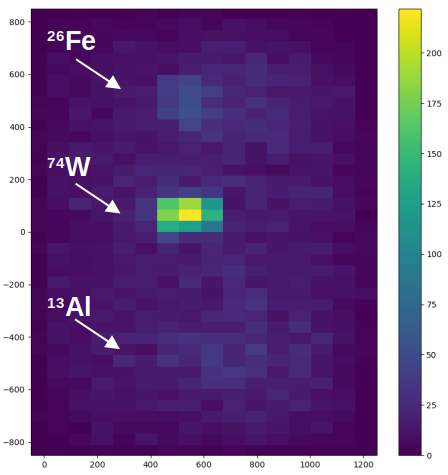}
\caption{Horizontal projection of PoCA points restricted to events with angular deflection above 11$\degree$. The tungsten target of rectangular shape ($20\times10\times5$ cm$^3$ ($x \times y \times z$)) can easily be identified at the center of the image.}
\label{poca_horizontalProj}
\end{figure}

\begin{figure}[ht] %[htbp]
\centering
\includegraphics[width=.48\textwidth, trim = 0 0 0 0, clip=true, angle=0]{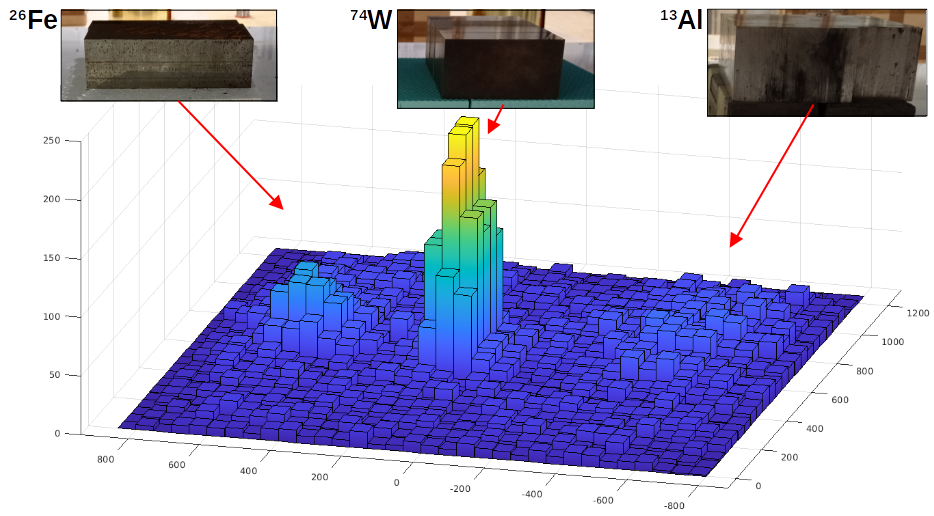}
\caption{2D projection of PoCA points restricted to events with angular deflection above 10$\degree$.}
\label{poca_3DProj}
\end{figure}

Approximately $1.1 \times 10^6$ events were recorded during 5 hours of acquisition, with targets of aluminium, iron and tungsten at the center of the telescope, as shown in figure~\ref{Al_Fe_W_test}. The horizontal projection of the PoCA points, while requiring angular deflections above 10$\degree$-11$\degree$, can be seen in figure~\ref{poca_horizontalProj} and the respective 3D projection is given in figure~\ref{poca_3DProj}.

The low angular resolution of the telescope, due to the small distance between RPCs (45 cm) and spatial resolutions of the order of the centimeter, explains the high angular restriction (above {\small$\sim$}10$\degree$) imposed to the PoCA points. To be able to detect scattering angles around 1$\degree$, corresponding to the typical rms deflection produced by 10 cm of tungsten, detectors with similar spatial resolution should have been placed 10 times further away, of course hardly feasible. On the other hand, an improvement of the detector's spatial resolution, also by a factor of 10, would have the same impact on the angular resolution.

The applied angular restriction allowed us to maximize the Signal-to-Noise Ratio (SNR), as clearly observed experimentally, however at the price of discarding a significant amount of good events. In fact, as a consequence of the mentioned angular restriction, only 1\% of the total number of events were used in figures~\ref{poca_horizontalProj} and~\ref{poca_3DProj}.
%only 1\% of the total number of events were available to perform figures~\ref{poca_horizontalProj} and~\ref{poca_3DProj}.

\begin{figure}[ht] %[htbp]
\centering
\includegraphics[width=.48\textwidth, trim = 0 0 0 0, clip=true, angle=0]{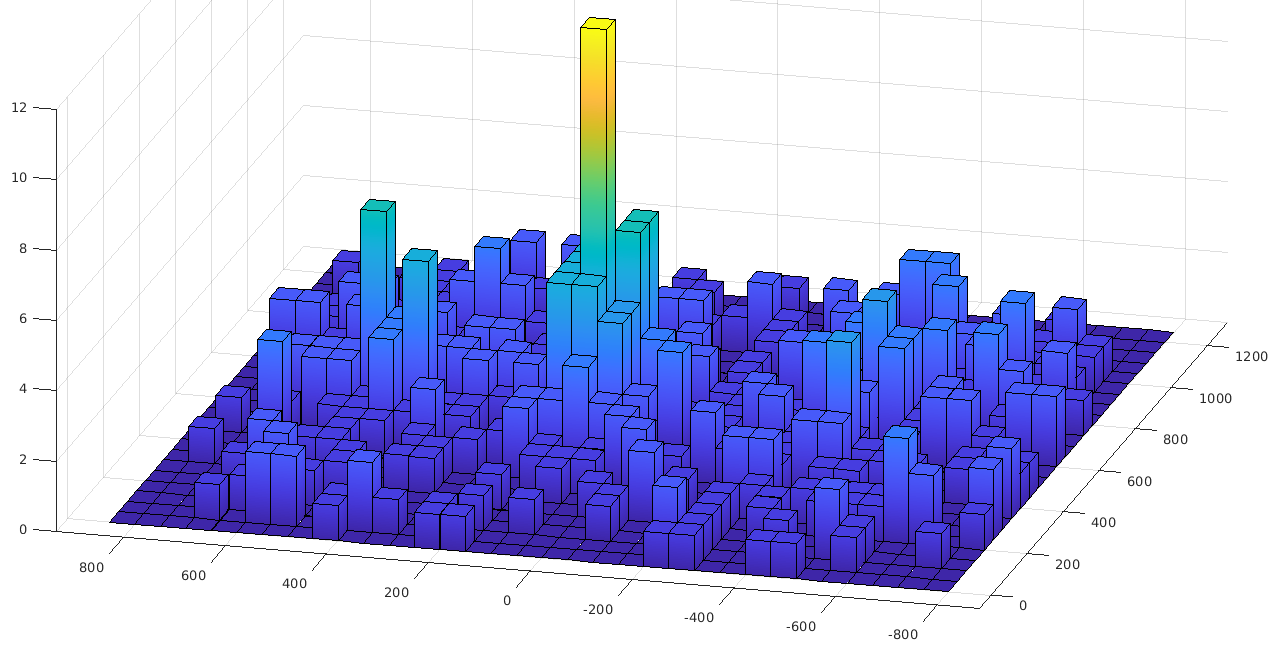}
\caption{3D projection of PoCA points restricted to events with angular deflection above 10$\degree$ and obtained with only 10 minutes of acquisition.}
\label{poca_3DProj10mins}
\end{figure}

Nevertheless, despite the current spatial resolution and short distance between planes, the telescope needed only 10 minutes of acquisition in order to identify the presence of the 5 cm thick tungsten block, as shown in figure~\ref{poca_3DProj10mins}.

\section{Conclusion}

A muon telescope equipped with four Resistive Plate Chambers (RPC) of 2 m$^{2}$ per plane and spaced by 45 cm was tested during several hours with aluminium, iron and tungsten blocks of 5 cm of thickness, located at the center of the telescope. After only 10 minutes of acquisition, it was already possible to discrimate the tungsten block, and after several hours, the iron and tungsten blocks were clearly visible.

These RPCs were designed to work outside the traditional laboratory environment and at very low gas flow regime. The results here presented indicate that, at relatively low cost, an RPC based telescope could be used in a real application, identifying the presence of high Z materials in large volumes. Moreover, improvements regarding the current setup could be achieved by: (1) increasing the distance between detector planes, necessary in any case when imaging large volumes; and (2) increasing the position resolution of the RPCs, which would also significantly reduce the acquisition time.
%, for instance, one order of magnitude would open the possibility to reach acquisition times below the minute. 

\section{Acknowledgements}
This work was supported by Fundação para a Ciência e Tecnologia, Portugal in the framework of the project CERN/FIS-INS/0006/2021.

%\section*{References}
%\bibliography{mybibfile}

\begin{thebibliography}{99} % Use for 10-99 references

\bibitem{PDG_2020}
PDG, Review of Particle Physics, Progress of Theoretical and Experimental Physics 083C01, 2020. doi: 10.1093/ptep/ptaa104

\bibitem{Borozdin_2003}
K. Borozdin et al., Radiographic imaging with cosmic-ray muons. Nature 422, 277, 2003.
doi: 10.1038/422277a

\bibitem{Riggi_2013}
S. Riggi et al., Muon tomography imaging algorithms for nuclear threat detection inside large volume containers with the Muon Portal detector. NIMA 728, 59-68, 2013.
doi: 10.1016/j.nima.2013.06.040

\bibitem{Neiser_2013}
A. Neiser et al., TRB3: a 264 channel high precision TDC platform and its applications. JINST 8 C12043, 2013.
doi: 10.1088/1748-0221/8/12/C12043

\bibitem{Saraiva_2022}
J. Saraiva et al., Advances towards a large-area, ultra-low-gas consumption RPC detector. NIMA 1046, 167744, 2022.
doi: 10.1016/j.nima.2022.167744

\end{thebibliography}

\end{document}